# Domain structure of epitaxial Co films with perpendicular anisotropy

J. Brandenburg[a], R. Hühne, L. Schultz, V. Neu[b]

*IFW Dresden, Institute for Metallic Materials, P.O. Box 270116, 01171 Dresden, Germany, and*

*Institute for Solid State Physics, University of Technology Dresden, 01062 Dresden, Germany*

## Abstract

Epitaxial hcp Cobalt films with pronounced c-axis texture have been prepared by pulsed lased deposition (PLD) either directly onto $Al_2O_3$ (0001) single crystal substrates or with an intermediate Ruthenium buffer layer. The crystal structure and epitaxial growth relation was studied by XRD, pole figure measurements and reciprocal space mapping. Detailed VSM analysis shows that the perpendicular anisotropy of these highly textured Co films reaches the magnetocrystalline anisotropy of hcp-Co single crystal material. Films were prepared with thickness t of 20 nm < t < 100 nm to study the crossover from in-plane magnetization to out-of-plane magnetization in detail. The analysis of the periodic domain pattern observed by magnetic force microscopy allows to determine the critical minimum thickness below which the domains adopt a pure in-plane orientation. Above the critical thickness the width of the stripe domains is evaluated as a function of the film thickness and compared with domain theory. Especially the discrepancies at smallest film thicknesses show that the system is in an intermediate state between in-plane and out-of-plane domains, which is not described by existing analytical domain models.

---

[a] Present address: Max Plank Institute for Chemical Physics of Solids, Nötnitzer Str. 40, 01187 Dresden, Germany
[b] Corresponding author. Email: v.neu@ifw-dresden.de

## I.     Introduction

Thin magnetic films with perpendicular anisotropy recently received a renewed scientific interest due to the change from in-plane to perpendicular recording media in the magnetic data storage technology. For this application, the effective uniaxial anisotropy of the material and the orientation of its easy magnetization axis perpendicular to the surface has to be large enough to overcome the natural demagnetizing effect arising from the shape anisotropy of a thin film. Polycrystalline Co films typically possess a pure in-plane magnetization as the uniaxial magnetocrystalline anisotropy averages out and the domain structure is dominated by the large demagnetizing energy of $K_d = (1/2\mu_0)J_s^2 = 1.35$ MJ/m$^3$. However, if prepared as a single crystal film and with perpendicular orientation of the c-axis, the uniaxial anisotropy of hcp-Co ($K_u = 0.45$ MJ/m$^3$) can compensate at least partially the in-plane anisotropy. From films with such competing anisotropies a non trivial domain structure is expected. Vice versa, the study of the domain structure is a useful approach to examine the participating energy terms present in the material. In films with large perpendicular anisotropy domain theory and experimental observation often lead to good qualitative and quantitative agreement, as demonstrated e.g. for L1$_0$-ordered FePt films[1] or epitaxial Nd$_2$Fe$_{14}$B films.[2] Reliable domain interpretation in Co films with competing shape and magnetocrystalline anisotropy is, however, complicated by the limited applicability of existing domain theories.[3] The focus of this work is on thin, epitaxially grown Co films with precise control of the crystal structure and magnetocrystalline anisotropy, where the epitaxial relation to the substrate guarantees a perpendicular orientation of the c-axis with respect to the film plane. For such films a transition from pure in-plane domains into stripe domains with perpendicular magnetization component is expected. In extension to the few works which report on similar samples,[3,4] the thickness range from 20 to 50 nm has been studied in detail to determine the crossover thickness and domain width for above mentioned transition. Furthermore in all samples the effective uniaxial anisotropy of single crystal Co was proven by an analysis of the global magnetization process. In these well characterized films the evolution of the domain structure with film thickness can be critically compared with prediction from domain theory.

## II. Experimental

Thin Co films with varying thickness t (20 nm < t < 100 nm) were prepared on heated (100°C –500°C) single crystal $Al_2O_3$(0001) substrates by pulsed laser deposition (PLD) in a UHV chamber with a background pressure of $10^{-9}$ mbar. In few cases, an additional 5 to 10 nm thick Ru film was deposited as a buffer layer. Elementary targets (Co and Ru) were moved into the beam of an KrF excimer laser (248 nm, 25 ns, 3 J/cm$^2$) and the ablated material was deposited in direct on-axis geometry onto the substrate placed at a distance of 72 mm opposite to the target. Details on the imaging conditions of the laser beam and the ablation of Co can be found in Brandenburg *et al.*[5] The film thickness was adjusted by an appropriate choice of pulses onto the target and determined after the deposition by means of energy dispersive x-ray analysis (EDX). For these measurements, comparative EDX spectra of film and elemental bulk standards were analysed with a thin film software. Details are given in Neu *et al.*[6] The crystallinity of the films was measured by standard θ-2θ x-ray diffraction (XRD) in Bragg-Brentano geometry with a Philips X'Pert diffractometer and Co-$K_α$ radiation. Pole figures were measured with a Philips X'Pert texture goniometer and Cu-$K_α$ radiation to analyze the epitaxial growth of the individual layers. In case of the Ru buffer layer, the resolution of the standard pole figure goniometer is not sufficient to separate the reflections of the $Al_2O_3$ substrate and the buffer layer. Therefore, reciprocal space mapping was applied in a high resolution Philips X'Pert MRD goniometer using Cu-$K_α$ radiation, a Ge monochromator in the incoming beam and 0.09° soller slits on the detector side. The effective magnetic anisotropy of each film was determined by measuring the magnetic hysteresis with a Quantum Design PPMS vibrating sample magnetometer. For this, the area enclosed by the initial magnetization curves, recorded in in-plane and out-of-plane field orientation, was compared to the value obtained from an isotropic reference sample (prepared at room temperature on a Si substrate with amorphous SiN surface layer). The magnetic domain structure was imaged with a Digital Instruments Dimension 3100 scanning force microscope using high resolution MFM tips with single side coating (SC-20 by SmartTip B.V). The tip magnetization in the z-direction leads to a sensitivity to gradients in the z-component of magnetic stray fields above the sample surface.

### III. Results and Discussion

A. Epitaxial growth of Co(0001) on $Al_2O_3$

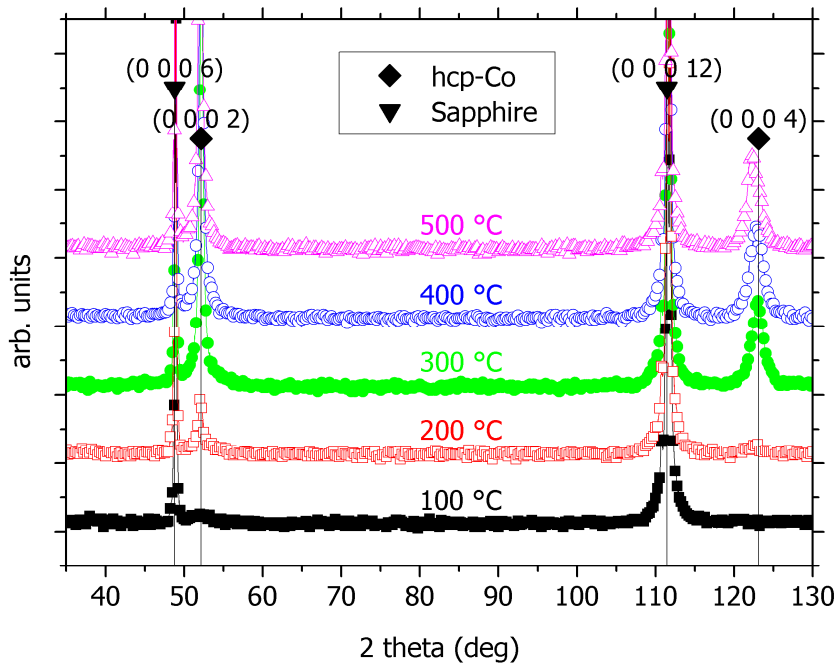

FIG. 1. (Color online) X-ray diffraction patters of Co films prepared at different substrate temperatures.

Fig. 1 shows the XRD data for about 80 nm thick Co films deposited at different substrate temperatures directly onto the sapphire substrate. For temperatures of 100°C and below only two sharp substrate reflections (Al$_2$O$_3$ (0 0 0 6) and (0 0 0 12)) are visible. At 200 °C two additional peaks evolve, which can be identified with the (0002) and (0004) reflection of the hcp-Co phase, demonstrating a textured growth with the c-axis perpendicular to the surface. A further increase of substrate temperature leads to an increased intensity of the Co reflections, which levels off at about 400 °C. At 500 °C a slight shift of the second Co peak is visible, which can be attributed to the growth of fcc rather than hcp Co, in agreement with the transformation temperature of 422 °C known from the equilibrium phase diagram of Co.

Pole figure measurements on the Co (1 0 –1 1) reflection (2θ$_{Cu}$ = 47.42°) in comparison with the Al$_2$O$_3$ (20-22) pole (2θ$_{Cu}$ = 46.17°) were performed on all films to determine the in-plane texture and a possible epitaxial relation to the substrate. Fig. 2a shows the result obtained on an 80 nm thick film deposited at 400 °C, where both measurements are overlayed graphically. Besides the expected sharp reflections of the sapphire single crystal with threefold symmetry, the signal of the Co reflection is seen as six evenly spaced discrete poles arranged on a ring corresponding to a 61° tilt away from the substrate normal. This is in agreement with the c-axis textured growth of the hexagonal Co grain and a unique epitaxial orientation within the basal plane. Furthermore a 30° rotation in the plane with respect to the substrate is observed. The full epitaxial relation is given as:

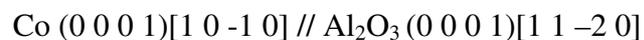
Co (0 0 0 1)[1 0 -1 0] // Al$_2$O$_3$ (0 0 0 1)[1 1 –2 0]

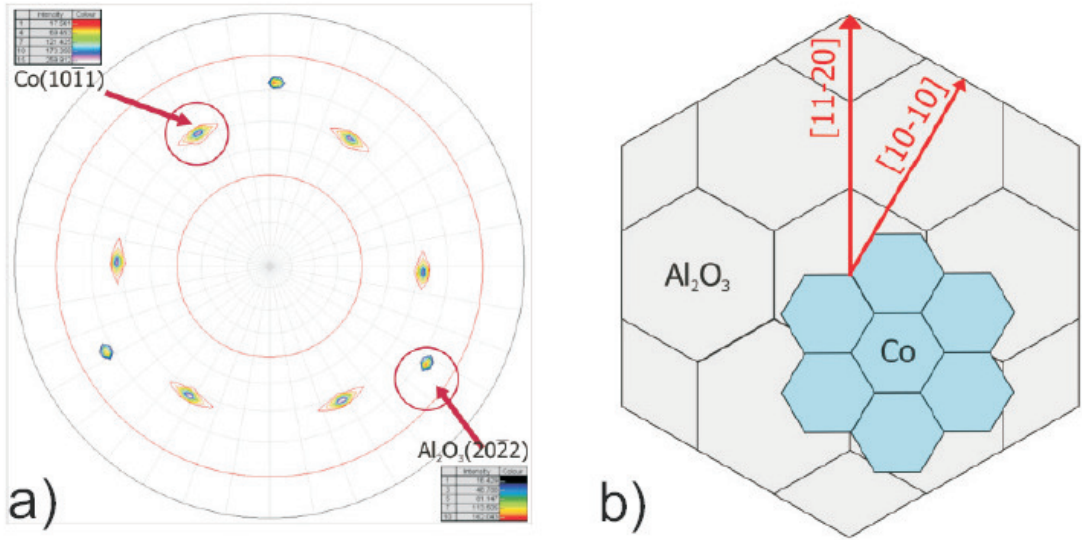

FIG 2. (Color online) (a) Pole figure measurements of $Al_2O_3$ and Co, and (b) sketch illustrating the epitaxial relation between film and substrate.

This growth mode holds for all films in which crystalline reflections can be observed, indicating, that the epitaxial relation dictates the grain orientation as soon as the substrate temperature is sufficiently large to result in a film with hexagonal crystal structure. The texture spread as determined from the full width at half maximum (FWHM) of the (1 0 -1 1) pole manifests a rather small tilt in out-of-plane direction of $\Delta\Psi = 3°$ and a rotational variation in-plane of $\Delta\Phi = 6°$. Fig. 2b illustrates again the orientation relation between sapphire substrate and Co film. Note that of course no lateral orientation configuration can be deduced from the x-ray experiments and the illustration is thus purely schematic. It shows, however, the size relation between the $Al_2O_3$ and Co unit cells and allows to estimate the lattice mismatch to be about -9%.[7]

The epitaxial growth has to be understood from the matching symmetry between the 6 fold $Al_2O_3$ surface and that of the hexagonal basal plane of hcp-Co, and from a tolerable lattice mismatch and justifies the choice of the $Al_2O_3$ (0001) substrate. Whether these conditions are decisive cannot be decided from this study. Interestingly, in other works reporting hcp-Co films with perpendicular c-axis orientation the films were grown on either $Al_2O_3$ (11-20) [8] or onto rigid mica substrates,[4] however a Ru (0001) buffer was utilized to support the desired c-axis texture.

B. Epitaxial growth of Co(0001) on Ru(0001)/$Al_2O_3$(0001)

To investigate the role of a Ru buffer layer on the epitaxial growth of Co on sapphire, a 5 to 10 nm thick Ru film was grown at 500 °C onto $Al_2O_3$ (0001), before the Co layer

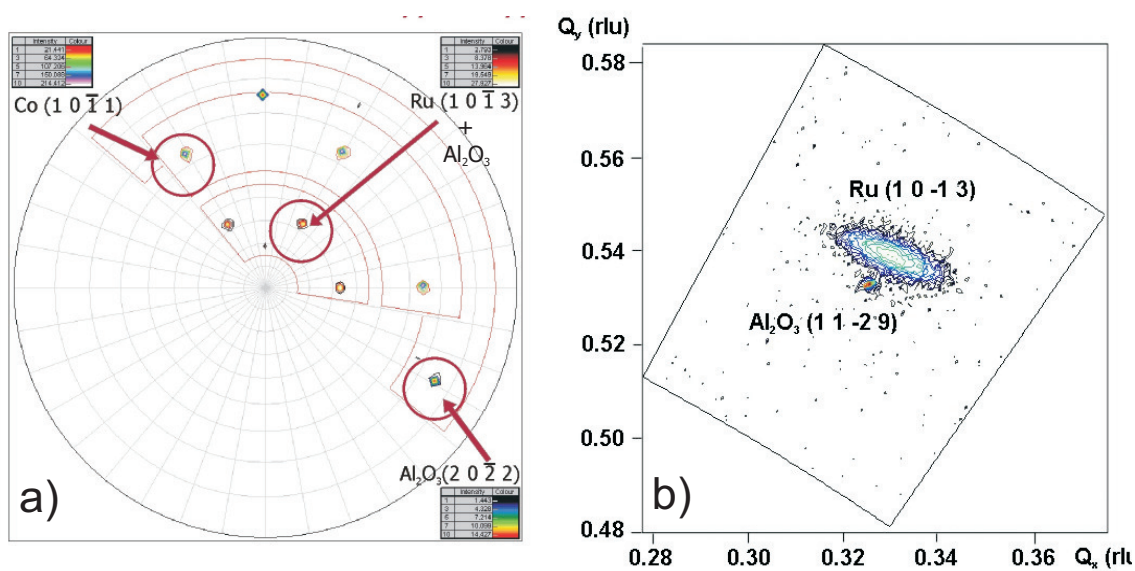

FIG 3. (Color online) (a) Pole figure measurement, (b) XRD reciprocal space map on a logarithmic intesity scale of the $Al_2O_3$ (1 1 -2 9) reflection at $Q_x$ = 0.325 rlu, $Q_y$ = 0.5324 rlu with the neighboring Ru(1 0 -1 3) pole at $Q_x$ = 0.330 rlu, $Q_y$ = 0.5385 rlu. The reciprocal length unit rlu is defined as $2\lambda/d$.

was deposited at 400°C. Judged from a work by Yamada *et al.*,[9] where Ru was deposited by sputtering at 500 °C on $Al_2O_3$ (0001), an epitaxial growth with c-axis texture and 30° in-plane rotation is expected. Fig. 3a shows a pole figure measurement of a Co/Ru/$Al_2O_3$ sample, where the Co (1 0 –1 1) pole, the Ru (1 0 –1 3) pole ($2\theta_{Cu}$ = 78.55) and the $Al_2O_3$ (2 0 –2 2) pole are overlaid graphically. The identical 6 discrete poles measured at 2θ = 78.55° were identified by Yamada *et al.* in their experiment as Ru reflections, as no 6-folded symmetry was expected from the rhombohedral substrate. However, although a measurement at fixed 2θ-value of 78.55° results in above pole figure, a θ-2θ scan on these poles results in a peak at 2θ = 77.2°, not explainable by hexagonal Ru, but by the (1 1 –2 9) $Al_2O_3$ reflection, which, as a substrate reflection still contributes to the x-ray intensity at 2θ= 78.55°. This explanation was easily proven by measuring a pole figure at 2θ = 77.2° on an uncovered substrate, which lead to an identical pole figure as shown in Fig. 3a. It is clear, that the experiment of Yamada *et al.* cannot verify the claimed epitaxial relation, but it is nevertheless a possible growth mode. Therefore, reciprocal space mapping was performed using a high resolution diffractometer in order to study the epitaxial relationship between Ru and the sapphire substrate in detail. The $Al_2O_3$ (1 1 –2 9) peak having a 2θ value of 77.22° (Cu-$K_\alpha$ radiation) was chosen for these investigations. The result of the measurement is shown in Fig. 3b. The sharp peak with very high intensities corresponds to the $Al_2O_3$ (1 1 –2 9) substrate reflection whereas the broader peak with low intensity was identified as Ru (1 0 –1 3). Additional measurements using different in-plane orientations of the substrate revealed that the Ru peak has a 6-fold symmetry similar to the $Al_2O_3$ peak, and thus the epitaxial growth relation of Ru on $Al_2O_3$ is determined correctly as:

Ru(0 0 0 1)[1 0 -1 0]  //  $Al_2O_3$(0 0 0 1)[1 1 –2 0]

The lattice mismatch between Ru and sapphire is –1.5%.[10] The subsequent Co film now grows in a 1:1 fashion onto the Ru buffer, as can be seen in Fig. 3a from comparing the (hidden) Ru (1 0 –1 3) poles with Co (1 0 –1 1). The epitaxial orientation towards the substrate is thus identical to that observed on the non-buffered sapphire substrate. The mismatch is slightly reduced to –8% and the overall epitaxial relation reads:

Co (0 0 0 1)[1 0 –1 0]  //  Ru(0 0 0 1)[1 0 -1 0]  //  $Al_2O_3$(0 0 0 1)[1 1 –2 0]

The c-axis texture and magnetocrystalline anisotropy (see next section) of Co/Ru/Al$_2$O$_3$ is not improved over that of Co films grown directly on Al$_2$O$_3$. Therefore, in the following only simple Co/Al$_2$O$_3$ films were investigated. A suitable hexagonal buffer, which mediates c-axis growth on sapphire, may however become of interest for the deposition of more oxygen sensitive films, e.g. Rare-earth –Co intermetallics, which need to be protected with a diffusion layer.[11]

C. Effective uniaxial anisotropy

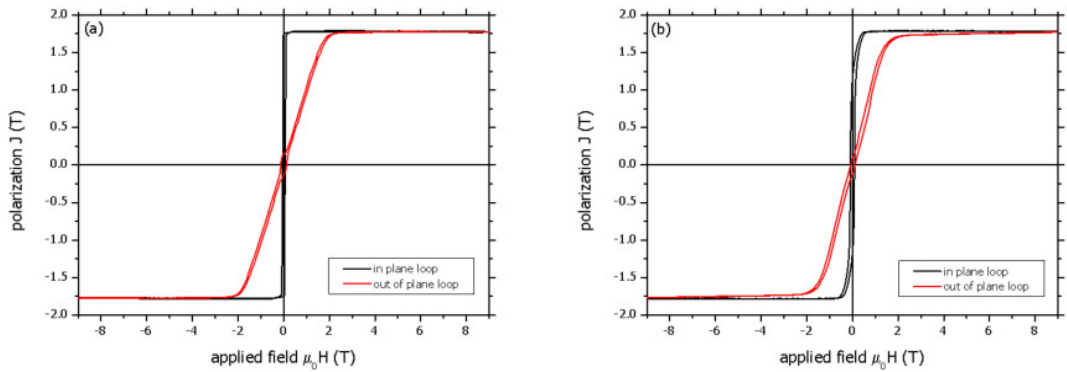

FIG 4. (Color online) Hysteresis loops of two 40 nm thick epitaxial Co films prepared at (a) room temperature and (b) 400°C showing strong anisotropy when measured in two orthogonal directions (in plane and out of plane).

Fig. 4 displays the magnetic hysteresis of two epitaxial Co films prepared at room temperature and 400°C substrate temperature, respectively. In both cases the magnetization shows only very little hysteresis but a large qualitative difference when measuring either with the field parallel (in-plane: ip) or perpendicular (out-of-plane: oop) to the film plane. In both samples the in-plane direction leads to a steeper hysteresis and a larger absolute polarization over the whole field range, despite the perpendicular uniaxial magnetocrystalline anisotropy of the epitaxially growing Co grains. Thus the shape anisotropy dominates the overall anisotropy of the sample. However, this effective in-plane anisotropy reduces when films are grown at higher temperature, as can be qualitatively deduced from the reduced saturation field in oop direction (Fig. 4b). Due to the competing anisotropies none of the hysteresis measurements are pure easy axis or hard axis procesess and the anisotropy can not be estimated simply from that saturation field. Torque measurements performed in saturation would lead to purely reversible rotational magnetization processes and are therefore the recommended choice for the analysis of anisotropies in such a case, but were not possible with the available experimental equipment. Therefore an alternative integral method was used, which was tested to give comparable values with a 3% precision.[12] For that, initial curves in the first quadrant of a magnetization loop are measured in both, ip and oop direction (Fig. 5a) and the enclosed integral area is determined. A fully isotropic film (grown at room temperature on an amorphous buffer) serves as a reference sample for a purely shape anisotropy driven behavior.

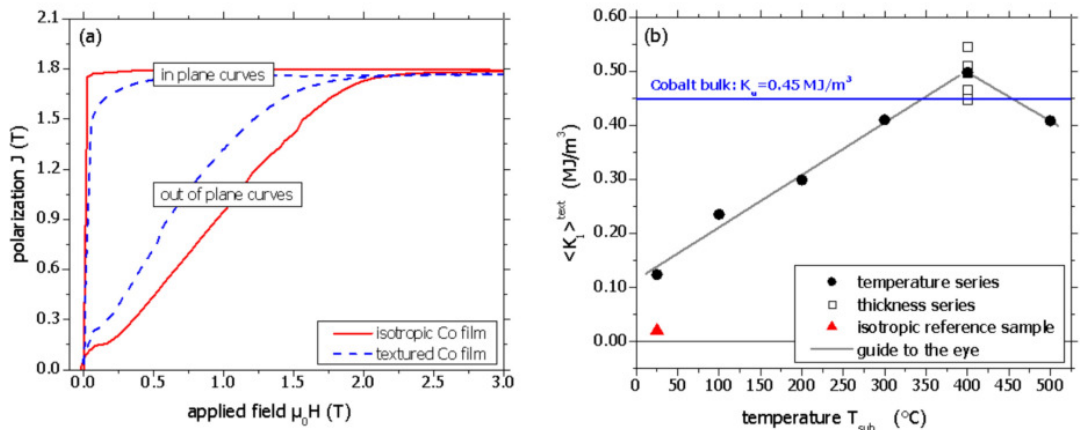

FIG 5. (Color online) (a) First quadrant initial loops of a 40 nm thick textured Co film in comparison to an isotropic reference film; (b) summary of the effective magnetocrystalline anisotropy as a function of deposition temperature.

When defining the above mentioned area as $\int(J_{ip} - J_{oop})dH$, the shape anisotropy contributes positively to this value, whereas the perpendicular anisotropy arising from the magnetocrystalline contribution of the textured Co-grains counts negatively. The area enclosed by the textured film is thus reduced over that of the isotropic sample. By subtracting the (in-plane) shape anisotropy (i.e. the demagnetizing energy $K_d$) the remaining perpendicular anisotropy can be evaluated. Fig. 5b shows the values for roughly 40 nm thick Co films prepared at different substrate temperatures from RT to 500°C (full circles), for films with thickness between 20 nm to 100 nm, all prepared at a substrate temperature of 400°C (open squares), and for the isotropic reference sample (triangle). In all cases, the such determined perpendicular anisotropy depends on both, the uniaxial anisotropy constant $K_1$ of the individual grains and the c-axis distribution function of the whole grain ensemble in the film and is thus also denoted as the texture averaged anisotropy constant $<K_1>^{text}$. As expected, $<K_1>^{text}$ of the isotropic reverence sample is close to zero. For films prepared on sapphire substrate $<K_1>^{text}$ increases steadily with the substrate temperature and reaches a maximum of about 0.45 – 0.5 MJ/m$^3$ at a deposition temperature of 400°C. As in the pole figure measurements no significant change of the c-axis spread with substrate temperature was observed, the increasing $<K_1>^{text}$ must be largely due to an improved uniaxial anisotropy $K_1$ of the individual Co grains. The maximum at 400°C reaches the anisotropy value for hexagonal single crystal Co, thus the anisotropy of the Co grains in the epitaxial film is fully developed. Comparably high values are furthermore obtained for all films within the thickness series (open squares), so that in the later interpretation of the domain structure of these films a uniaxial anisotropy constant of single crystal hcp-Co can be assumed, irrespective of the film thickness. No systematic dependency of $<K_1>^{text}$ on the film thickness is observed, indicating that surface anisotropy effects do not yet play a role. This is in agreement with an estimated surface contribution of less than 20% (10%) for films thicker than 10 nm (20 nm). For surface anisotropy contributions see C. Chappert *et al.*[13] The 40 nm thick film prepared at a substrate temperature of 500 °C has a $<K_1>^{text}$ value reduced over that of single crystal Co. This is explained by the occurrence of the fcc-Co phase observed in the x-ray investigation.

D. Domain structure

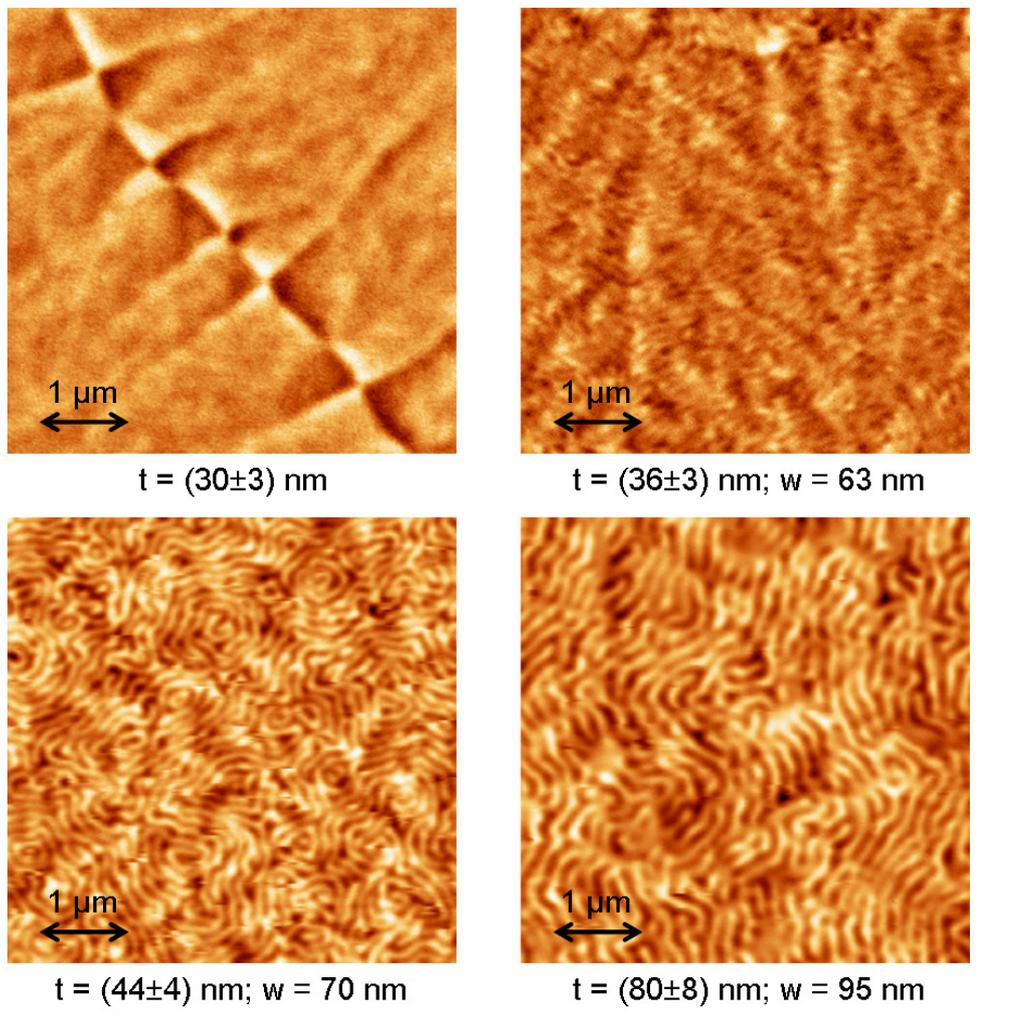

FIG 6. (Color online) MFM images of four films with 30, 36, 44 and 80 nm thickness. For films of 36 nm and above clear perpendicular MFM contrast on a length scale below 100 nm is observed. The film with 30 nm thickness shows a cross tie walls – a clear indication of in-plane magnetization (see text).

The magnetic domain structure of four exemplary Co films with thickness 30 nm, 36 nm, 44 nm, and 80 nm is seen in Fig. 6. For films of 36 nm and above a pattern of stripes with alternating dark and bright contrast is observed. The stripe width varies from 60 to 100 nm. They run roughly parallel on a length scale of some micrometer, but their orientation is randomly distributed on the larger scale of the 10 µm x 10 µm sized image. The contrast can be interpreted as coming from micron sized domains with a predominant in-plane component due to the shape anisotropy, which are oriented isotropically in all directions. The perpendicular magnetocrystalline anisotropy of the epitaxially grown Co grains leads to an additional perpendicular magnetization component, which alternates on a smaller length scale and gives rise to the observed stripe pattern. Such domains have been suggested by Saito *et al.*[14] for thin Ni-Fe films with (shape dominated) in-plane magnetization and a small additional strain induced perpendicular anisotropy. In films with smaller thickness no stripe domains can be observed. As the lateral resolution of the MFM measurement is by far not reached for the film with the smallest imaged domain width, the vanishing domain contrast has to be interpreted as a dying out of the perpendicular magnetization component. Instead, features with a length of some micrometers are seen in the MFM image which are correlated with topography and are thought to arise from the domain boundaries of the mentioned in-plane domains. In some areas cross-tie walls are observed (Fig. 6, t = 30 nm), which are known in soft magnetic thin films with in-plane magnetization and arise from a sequence of 90° Néel walls to avoid energetically more costly 180° Néel walls. By examining a large number of films in the thickness range between 20 nm and 50 nm the critical thickness for stripe nucleation is found as $t_{cr}$= (36 ± 3) nm, with the error being determined by the precision of the thickness measurement. The average stripe domain width for all films with thickness 36 nm and larger is summarized in Fig. 7. Thinner films, which do not develop a stripe domain pattern, are listed with an arbitrarily chosen domain width of zero to demonstrate the sharpness of the transition. Within the accuracy of the measurement the domain width increases monotonously from about 60 nm to 95 nm. (red squares). Included in Fig. 7 are the data by Hehn[3] (black triangles), which cover the film thickness range from 50 nm to 150 nm. In the overlapping thickness range the domain width data compare well.

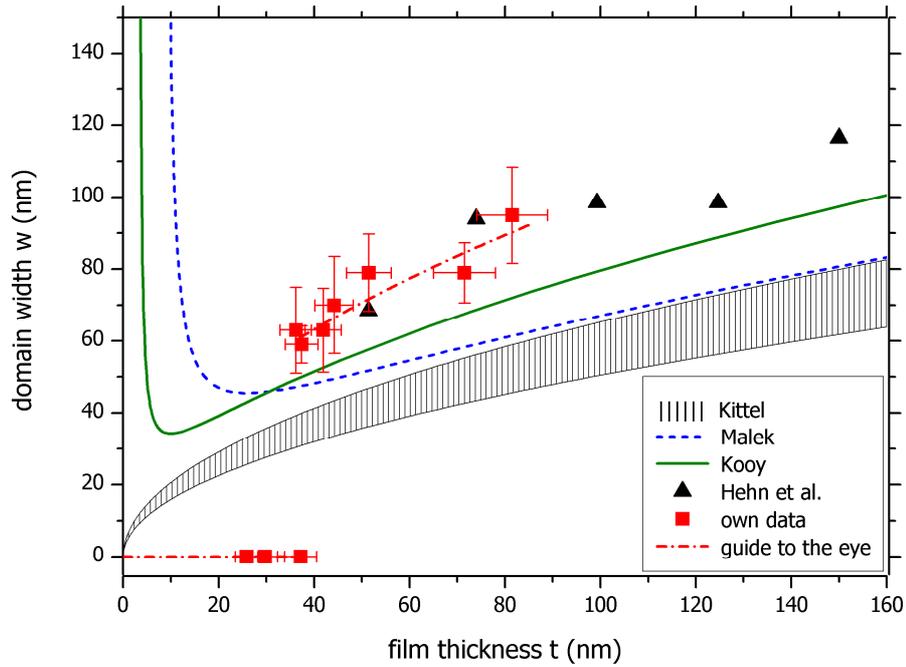

FIG 7. (Color online) Average stripe domain width as a function of the film thickness in comparison with predictions from analytical domain models. Films which do not develop stripe domains are displayed with an arbitrarily set domain width of zero.

A good understanding of the domain structure is achieved if the experimentally observed thickness dependency can be explained by domain theory, i.e. by a prediction of an average domain width as a function of the film thickness, assuming a simplified, but realistic domain structure, that considers the materials parameters of the investigated magnetic thin film and carries the characteristic features of the domain pattern. Furthermore, such a model domain structure has to be energetically favored over other pos-sible domain configurations. The stripe domain model by Saito *et al.* considers a case of competing anisotropies very similar to that of the epitaxial Co films. It predicts a pure in-plane magnetization for very thin films and the occurrence (nucleation) of the already mentioned stripes above a critical film thickness. Critical thickness and the domain width at the stripe nucleation depend on the materials parameters, the saturation polarization $J_s$, the perpendicular anisotropy constant $K_u$ and the exchange constant A. For $J_s$ = 1.8 T, $K_u$ = 0.45 MJ/m$^3$ and A = 28·10$^{-12}$ J/m critical thickness and domain width are calculated to be $t_{cr}$ = 37 nm and $w_{cr}$ = 52 nm, respectively. The predicted onset (critical thickness) of stripe nucleation agrees very well with the experimental behavior. The observed domain width is about 20% too large. A further comparison is, however, not possible, as at present no theory of the stripe domain width as a function of the film thickness is available. So far, only few cases have been treated numerically[15,16] and found an increased domain width with increasing film thickness. It is thus desirable to test simpler models and evaluate, to which extend they can be applied to the present situation. A domain model for parallel stripes with alternating perpendicular magnetization has been developed by Kittel,[17] which is valid under the assumption of a negligible domain wall width, a pure perpendicular magnetization, and a film thickness larger than the domain width.[18] Although its application to a domain structure with strong in-plane component is doubtful, the square root functional dependency of domain width of the film thickness was used in Hehn et al.[3] to compare it with their experimental data. Good agreement was found with a value of the exchange stiffness constant A = 85·10$^{-12}$ J/m, but neither is this value reasonable nor is an agreement between model calculations and experiment satisfactory, if essential characteristics of the model are not met by the reality. If domain width calculations based on the Kittel model are performed with a realistic exchange stiffness constant A = (10 – 28)·10$^{-12}$ J/m (hatched area in Fig. 7), it becomes obvious, that the experimental behavior can not be described. A more realistic model has to include the in-plane component present in the domain structure. The calculations by Kittel for perpendicular stripes have been modified by Malek and Kambersky[19] to cover also films with thickness t comparable to the domain width w, and by Kooy and Enz,[20] to describe films which possess an additional in-plane component in the magnetization. In the latter ansatz, a ratio $Q = K_u/K_d < 1$ between perpendicular uniaxial anisotropy and shape anisotropy, stabilizes a stripe domain structure, where the magnetization in each domain is tilted away from the perfect perpendicular orientation expected in films with $Q \geq 1$. Consequently, the equilibrium domain width of such a magnetization structure possesses a modified dependency on the film thickness. The dashed curve in Fig. 7 is a calculation after Malek and Kambersky with $J_s = 1.8$ T, $K_u = 0.45$ MJ/m$^3$ and $A = 28 \cdot 10^{-12}$ J/m. For large film thickness it approaches the square root dependency of the Kittel model, as expected. For smaller film thicknesses the refined model leads to increasingly smaller domain values with decreasing thickness and but also predicts a strong upturn in the equilibrium domain width for film thicknesses below 25 nm. When the shape anisotropy is considered by performing the calculations after Kooy and Enz with the experimentally given quality factor $Q = 0.35$ the domain width function modifies again (solid line). The minimum domain width is reduced and it occurs at a lower film thickness value. The increase in domain width with increasing film thickness is, however, steeper as in the pure perpendicular case. For $t = 150$ nm a 20% larger domain width is expected. Including an in-plane component in the predominant perpendicular domain structure thus leads to a more realistic description of the experimentally observed stripe domains. Nevertheless, the agreement is still far from being satisfactory. Qualitatively, in the stripe domain model by Saito *et al.* and in the experimental observation, at low film thickness the stray field minimization leads to a pure in-plane domain structure, whereas by insisting on a perpendicular magnetization component such as in [19] and [20] the domains simply grow larger. Quantitatively, in the region where stripe domains are observed, their domain width is still larger than predicted by Kooy and Enz. Possible reasons for the discrepancy are deviations of the experimentally observed domain width from the equilibrium width of long parallel stripes considered in the models. Such differences in the domain width of maze-like stripe domains and parallel stripe domains are e.g. reported in Co/Pt multilayers with strong perpendicular anisotropy,[21] but are not found in the Co films studied by Hehn et al.[3] A more serious reason for the discrepancy between experiment and model is thought to be the assumption of a homogeneous magnetization state within each domain and a vanishing domain wall width. Improvements over such simplifying assumptions are so far only attempted by micromagnetic methods.

Micromagnetic simulations of stripe domains in thin films have only been performed for few cases, such as Co-Pt alloy films[15] and most recent for Co films.[16] In that numerical study, stripe domain nucleation and the evolution of domain width with film thickness was calculated based on an energy minimization scheme (OOMMF) for a fixed set of Co material parameters ($J_s$ = 1.78 T, $K_u$ = 0.46 MJ/m$^3$ and A = 13·10$^{-12}$ J/m). They find a critical thickness for stripe nucleation of $t_{cr}$ = 22 nm, which is smaller than the experimentally observed value of 36 nm and the value calculated after Saito *et al.* ($t_{cr}$ = 37 nm).[14] At $t_{cr}$, the numerically obtained stripe domain width is $w_{cr}$ = 35 nm and above this critical thickness the equilibrium domain width is observed to closely follow a square root dependency as a function of film thickness, which is quantified as $w = 7.5 \cdot \text{nm}^{1/2} \cdot \sqrt{d(\text{nm})}$. This dependency nicely matches the experimentally observed increase in domain width concerning the functional dependency, but still leaves a discrepancy between the simulated and measured absolute values in domain width. Given the large spread in reported values for the exchange constant A in Co, better quantitative agreement might be achieved with slightly modified materials constants in the numerical simulations. The predictions from the different analytical models and the micromagnetic simulation are compared once more with the experimental observations in table I.

Table I: Critical thickness for stripe nucleation (if available) and equilibrium stripe domain width for film thicknesses of 36 nm and 80 nm as obtained from the experiment, the four analytical models and the micromagnetic simulation. In the evaluation of the analytical models the materials parameters were chosen as $K_u$ = 0.45 MJ/m$^3$, $J_s$ = 1.8 T and A = 28 x 10$^{-12}$ J/m. The numerical results were taken from reference 15 and were based on $K_u$=0.46 MJ/m$^3$, $J_s$ = 1.78 T and A= 13 x 10$^{-12}$ J/m.

|  | experiment | Saito | Kittel | Malek | Kooy | numerical |
|---|---|---|---|---|---|---|
| Critical thickness (nm) | 36 | 37 | - | - | - | 22 |
| Domain width (nm) (t = 36 nm) | 63 | 52 | 39 | 47 | 49 | 45 |
| Domain width (nm) (t = 80 nm) | 95 | - | 58 | 61 | 71 | 67 |

## IV. Conclusions

Co films have been grown epitaxially on pure and Ru-buffered Al$_2$O$_3$ (0001) substrates with hexagonal crystal structure and perpendicular alignment of the c-axis with respect to the film plane. The epitaxial growth of Ru on Al$_2$O$_3$ has been confirmed by high resolution reciprocal space mapping, and regular pole figure measurements revealed a single epitaxial growth mode for hcp-Co on both, pure and buffered substrate with a 30° rotation in the basal plane. The texture averaged perpendicular anisotropy reaches the magnetocrystalline anisotropy value of bulk hcp-Co, when the films are prepared at a substrate temperature of 400°C. Stripe domain nucleation is found for films with thickness of 36 nm and above, in agreement with the theory of Saito *et al*. Existing analytical domain models can only describe the functional dependency of the domain width on film thickness, but fail to reach satisfactory quantitative agreement. Still, including an in-plane tilt originating from the large demagnetizing energy, improves the perpendicular domain model after Kittel substantially. It is expected, that further micromagnetic simulations are required to fully describe the domain formation in this system with competing perpendicular magnetocrystalline anisotropy and in-plane shape anisotropy, which will eventually allow to deduce materials parameters from the evaluation of nanoscaled domain images.